Counter-rotating black holes from FRII lifetimes

David Garofalo

Department of Physics, Kennesaw State University, USA

Estimates suggest that while FRII jets appear to have lifetimes constrained to hundreds of millions of years, radio galaxies with FRI jets appear to be longer lived. We illustrate the nature of this time constraint from model perspectives, showing how compatibility between theory and data match in a way suggesting a key difference between active galaxies whose engines are characterized by accretion onto co-rotating versus counter-rotating black holes. We calculate a range of timescales for counter-rotating black holes for a range of accretion rates compatible with theory which we then compare to data. The validity of these timescales constitutes the most powerful recent piece of evidence for considering counter-rotation between black holes and accretion disks in high energy astrophysics.

## 1. Introduction

According to the current paradigm, powerful FRII jets in radio galaxies are the product of accretion onto rapidly spinning, prograde accreting black holes (Wilson Colbert 1995; Sikora et al 2007; Tchekhovskoy et al 2010; Tchekhovskoy & McKinney 2012). The FRII classification refers to jets that are more collimated and generally more powerful than FRI jets, the latter often being subjected to entrainment from the interstellar medium (Fanaroff & Riley 1974). Because prograde accretion onto a black hole can only spin black holes up, the only constraint for the longevity of the jet is the amount of accreting fuel. Radio galaxies can either be high excitation or low excitation, depending on the degree of emission line signatures or thermal versus non-thermal nature (Hine & Longair 1979; Best & Heckman 2012; Antonucci 2013; Mingo et al 2014; Macconi et al 2020; Mingo et al 2022). Because FRII radio galaxies are often low excitation systems, they should experience constraints on their lifetimes that are similar to those for FRI radio galaxies, such as M87, with timescales orders of magnitude longer than those associated with feeding at near Eddington rates. But this is not supported by the data. In fact, radio galaxies with FRII jets appear to have quantifiably limited timescales unlike their FRI counterparts (e.g. O'Dea et al 2009; Garofalo, Singh & Zack 2018). High excitation FRII systems, for example, are found to be limited to 10 million years (Turner & Shabala 2015). Most recently, Dabhade, Saikia & Mahato (2022) have compared radio galaxies with the giant radio galaxy population, including the lifetimes of jets in FRII sources, among others. This constitutes the most exhaustive quantitative analysis of FRII lifetimes and, if these results hold up to future scrutiny, a powerful constraint on the nature of jet formation and evolution in jetted active galactic nuclei (AGN). We suggest that the difference in measured timescales for powerful FRII jets compared to powerful FRI jets points to a basic difference in the nature of the two morphologies that was captured in the gap paradigm for black hole accretion and jet formation (Garofalo, Evans & Sambruna 2010). Whereas powerful FRII jets, in this paradigm, are produced in accreting black holes spinning in the opposite direction as the accretion disk (i.e. counter-rotation), the opposite is true for FRI jets. And since counter-rotation spins black holes down while co-rotation spins them up indefinitely, powerful FRII jets are limited in time in a way that powerful FRI jets are not. The possibility that radio galaxies with FRII jet morphology are constrained in time unlike FRI radio galaxies is, therefore, interesting in a fundamental way in high energy astrophysics. Evidence that FRII radio galaxies are constrained in time in a way that matches model predictions for the spin down timescales in both high and low excitation systems is thus exciting for understanding the nature of the longstanding puzzle behind the FRI/FRII jet dichotomy. In Section 2 we discuss the data analyzed in Dabhade et al (2022), describe the theory and emphasize the match between theory and data in Section 3. In Section 4 we conclude.

**2. Data**

Machalski, Koziel-Wierzbowska & Goyal 2021 explored the time evolution of 361 FRII radio galaxies from Cambridge, 3CRR, 6CE, 5C6, and 5C7 Sky Surveys and from the Bologna B2, Green Bank GB, and GB2 Surveys in order to produce a statistically relevant sample. They obtained a range of lifetimes for FRII radio galaxies which Dabhade et al 2022 plot in their Figure 6 and which we show on the left side of Figure 1. On the right-hand side of Figure 1 we show the maximum lifetime data from the left-hand side of Figure 1 on the dynamical age of the FRII jet with jet length obtained from Machalski et al 2021. In other words, we hone in on the 4 objects with oldest dynamical ages (circles) for each class of object, namely radio galaxies, radio quasars, and giant radio galaxies and giant radio quasars. The red objects represent high excitation FRII jetted AGN, i.e. with quasar or thermal-like signatures, indicative of radiatively efficient accretion. The red circle on the left side of the vertical line has the maximum dynamical age for an FRII high excitation radio galaxy or FRII HERG. The red object on the right side of the vertical line represents a giant radio quasar. I.e. it has all the same characteristics as the red counterpart on the left except for its jet length. The object on the right is considered a giant FRII HERG. The blue objects, similarly, distinguish themselves in the same way as the red objects do, except they belong to radiatively inefficient accretion, showing an absence of thermal or emission line signatures. They are thus FRII LERG for low excitation radio galaxies. As described in Section 3, we add theoretical values to Figure 1 with diamonds associating their theoretical age (i.e. the model prescribed duration of time for the FRII jet) with the same jet length values as the observational data for ease of comparison.

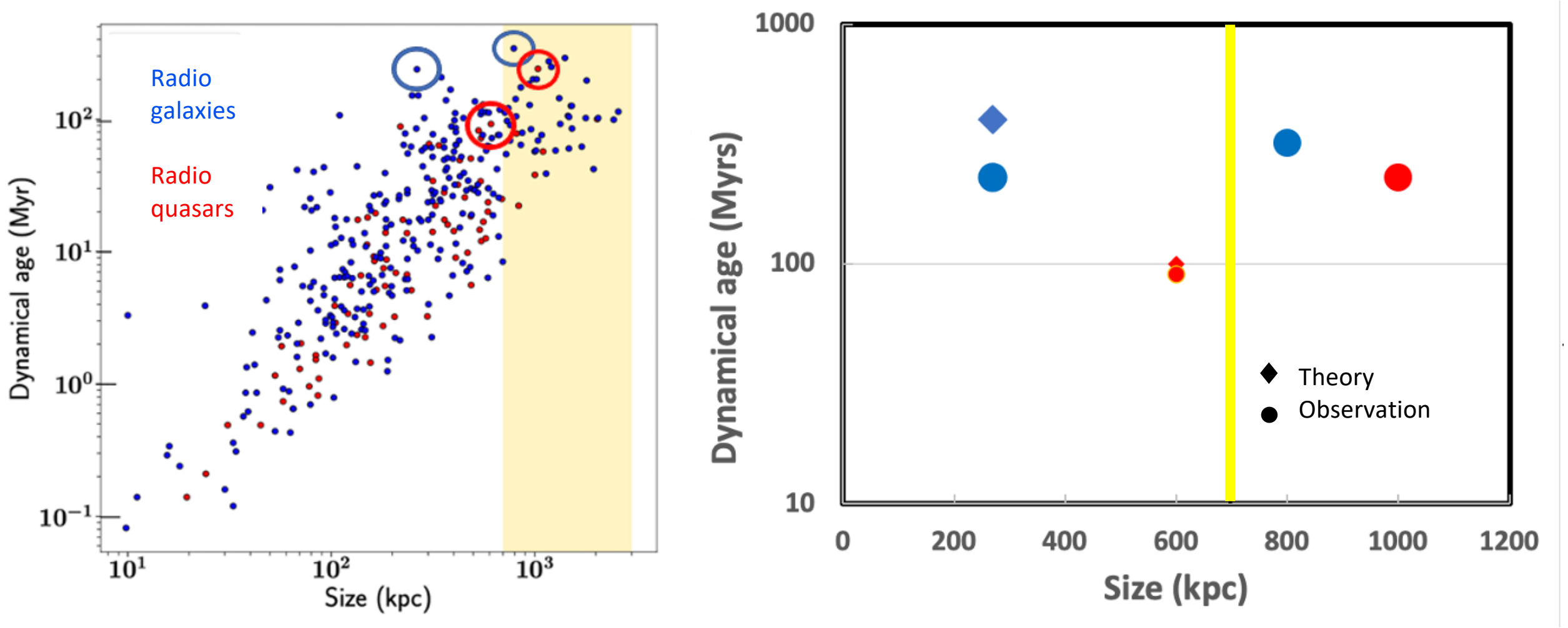

Figure 1. Left side: Dynamical age versus size of jet from Dabhade et al 2023. Radio quasars are in red and radio galaxies in blue. Radio galaxy (and giant radio galaxy) and radio quasar (and giant radio quasar) with maximum dynamical age indicated with appropriate blue and red circles, respectively, i.e. the four circular points on the right correspond to the objects at the centers of the red and blue circles on the left. Highlighted region in yellow is for giant radio quasars and giant radio galaxies. Right side: Maximum dynamical ages for radio quasars (red circles) with giant radio quasars on the right of the dividing line and for radio galaxies (blue circles) with giant radio galaxies on the right-hand side of the dividing line. The yellow dividing line represents the boundary between radio galaxies and giant radio galaxies. Data from Dabhade, Saikia & Mahato 2023. Theoretical values for FRII HERG and FRII LERG as described in the text are added (diamonds).

The maximum lifetime observed so far for FRII HERG and FRII LERG circled in red and blue on the left panel of Figure 1, with values equal to 9 x $10^7$ and under 3 x $10^8$ years, respectively, are compatible with estimates from theoretical modeling, which we will show in Section 3. While the lifetimes for giant radio quasars and giant radio galaxies increase from a theoretical perspective (Garofalo 2022), their value is more uncertain because such objects are not necessarily triggered at the Eddington accretion limit. Because of this uncertainty, we do not include theoretical maximum lifetimes for giant radio quasars and giant radio galaxies. Note that if a radio quasar had a lifetime equal to 230 million years (as the giant radio quasar indicated with the red circle), it would violate theory. The theoretical limit, in fact, is shown as the red diamond. For FRI radio galaxies, as mentioned above, no such time constraint is found. Two decades ago, evidence began to emerge suggesting that FRI jets live about an order of magnitude longer than FRII jets (Parma et al 2002). The evidence in this respect has grown (e.g. Saripalli et al 2012) and it was found that LERG systems live up to order $10^9$ years (Turner & Shabala 2015). The constraints on FRII jets from Machalski et al 2021 suggest, therefore, that long-lived LERG systems are FRI. In other words, an FRI radio galaxy can live substantially longer than FRII sources. Why is that?

## 3. Theory

In this section we describe how the timescales above emerge from, or are compatible with, theory. The theory is anchored to the idea that counter-rotation between black holes and accretion disks give rise to FRII jets (Garofalo, Evans & Sambruna 2010). As counter-rotation is an unstable and less likely configuration, the majority of mergers in this paradigm funnel cold gas into the nucleus that settles into co-rotation around a spinning black hole. It is therefore the minority that end up in counter-rotation and this is environment-dependent. It is also worth noting that an engine-based difference between FRII and FRI matters for the most powerful jets and that environment makes a difference at lower jet power (see Garofalo, Evans & Sambruna 2010 on the nature of the Owen-Ledlow diagram). We will therefore focus on the most powerful jets in the paradigm.

For the minority of configurations that end up in counter-rotating accretion disks around spinning black holes, we explore two basic evolutionary scenarios relevant for understanding the maximum possible lifetimes of FRII jets. The question we need to answer is this: How long does it take to spin a black hole down? This is because FRII jets are associated with counter-rotation. We want to find the maximum possible time for this process to then compare with the data for FRII lifetimes. Mergers yield initial conditions involving cold gas funneled into the galactic nucleus and the formation of a radiatively efficient disk accreting in counter-rotation at the Eddington limit. A subset of these counter-rotating black holes spin down at the Eddington limit while others spin down at accretion rates that begin at the Eddington limit but drop to rates as low as $10^{-2}$ the Eddington accretion rate. This range of accretion rates determines the range of jet lifetimes. We should point out that the initial spin value is crucial in determining jet lifetime. But since we are striving to determine maximum jet lifetimes, the initial spin value is assumed to be its theoretical maximum at 0.998. The drop in accretion rate depends on feedback processes, the details of which are not of present concern (see Garofalo, Evans & Sambruna 2010 for details). What matters here is the range of accretion for counter-rotation because it determines the lifetime for a counter-rotating black hole. This timescale depends on the amount of angular momentum added to the black hole by the accreted plasma and the rate at which the plasma accretes. The angular momentum accreted is that of the gas at the inner edge of the disk, which is referred to as the marginally stable circular orbit, $r_{ms}$. This radial location depends both on the spin of the black hole and the orientation of the accretion disk. From the stability of circular orbits in Kerr space-time one finds that $r_{ms}$ drops from 9 gravitational radii, $r_g$, for a maximally spinning black hole surrounded by a counter-rotating accretion disk, to just over 1 $r_g$ for a maximally spinning black hole surrounded by a co-rotating accretion disk (e.g. McClintock et al 2011). To determine the amount of angular momentum delivered to the black hole, one sets r to $r_{ms}$ in the following expressions for the angular momentum per unit mass as a function of Boyer-Lindquist radial coordinate r (Bardeen et al 1972),

$$L_+ = M^{1/2}(r^2 - 2aM^{1/2}r^{1/2} + a^2)/(r^{3/4}(r^{3/2} - 3Mr^{1/2} + 2aM^{1/2})^{1/2}) \quad (1)$$

and

$$L_- = - M^{1/2}(r^2 + 2aM^{1/2}r^{1/2} + a^2)/(r^{3/4}(r^{3/2} - 3Mr^{1/2} - 2aM^{1/2})^{1/2}), \quad (2)$$

where the '+' and '-' subscripts refer to the value of the angular momentum per unit mass as a function of radial coordinate for co-rotating disks and counter-rotating disks, respectively, with *a* the spin of the black hole and M its mass. By multiplying by accreted mass, one obtains the amount of angular momentum accreted, which changes the spin of the black hole. Because accretion rates are model-prescribed in terms of the Eddington value, our results are scale invariant, making the actual values of angular momentum irrelevant for our calculations. To see this scale invariance explicitly, we begin with the definition of the dimensionless black hole spin

$$a = c\,\mathcal{L}/(GM_{BH}^2) \tag{3}$$

where $\mathcal{L}$ is the angular momentum of the black hole and c is the speed of light. If accretion proceeds at the Eddington limit we set the Eddington luminosity to the luminosity in terms of the accretion rate and the disk efficiency as in equation (4). For our purposes we repackage all the constants into one term and relate the accretion rate to the black hole mass in equation (5), which can be written as in equation (6). The infinitesimal mass accreted onto the black hole is therefore obtained from equation (6) as shown in equation (7). We can carry out the analysis in the Newtonian limit with the magnitude of the angular momentum of the infinitesimal parcel of gas given to the black hole given in equation (8), with v the velocity of the parcel of gas dm at the inner edge of the disk that is supplied to the black hole and r is its radial location. From circular motion and Newton's 2nd law we obtain equation (9) from which we obtain the velocity in equation (10).

$$\eta\dot{M}c^2 = 4\pi c G m_p M_{BH}/\sigma\,. \tag{4}$$

$$\dot{M} = (\text{constant})\, M_{BH} \tag{5}$$

$$dM/dt = (\text{constant})\, M_{BH}. \tag{6}$$

$$dm = (\text{constant})\, M_{BH}\, dt. \tag{7}$$

$$d\mathcal{L} = dm\, v\, r \tag{8}$$

$$dm\, v^2/r = G\, dm\, M_{BH}/r^2 \tag{9}$$

$$v = (GM_{BH}/r)^{1/2}. \tag{10}$$

The inner edge of the disk depends on the black hole spin parameter and has the range given in equation (11). For our purposes we note that $r \propto GM_{BH}/c^2$ from which we get equation (12). For accretion at the Eddington limit, therefore, the rate at which the angular momentum of the black hole changes is shown in equation (13), from which we can determine how the dimensionless spin parameter of the black hole changes by using the differential form of equation (3) to obtain equation (14), from which we get equation (15), which is the black hole mass independent result we anticipated.

$$1.23GM_{BH}/c^2 \ - \ 9GM_{BH}/c^2 . \qquad (11)$$

$$d\mathcal{L} \propto \mathrm{dm}\ (GM_{BH}/r)^{1/2}\ GM_{BH}/c^2 \ \propto M_{BH}\ dt\ \mathrm{c}\ GM_{BH}/c^2 \ = M_{BH}\ dt\ GM_{BH}/c. \qquad (12)$$

$$d\mathcal{L}/dt \propto M_{BH}^2. \qquad (13)$$

$$da/dt = cd\mathcal{L}/dt/(GM_{BH}^2) \qquad (14)$$

$$da/dt \propto c/G \qquad (15)$$

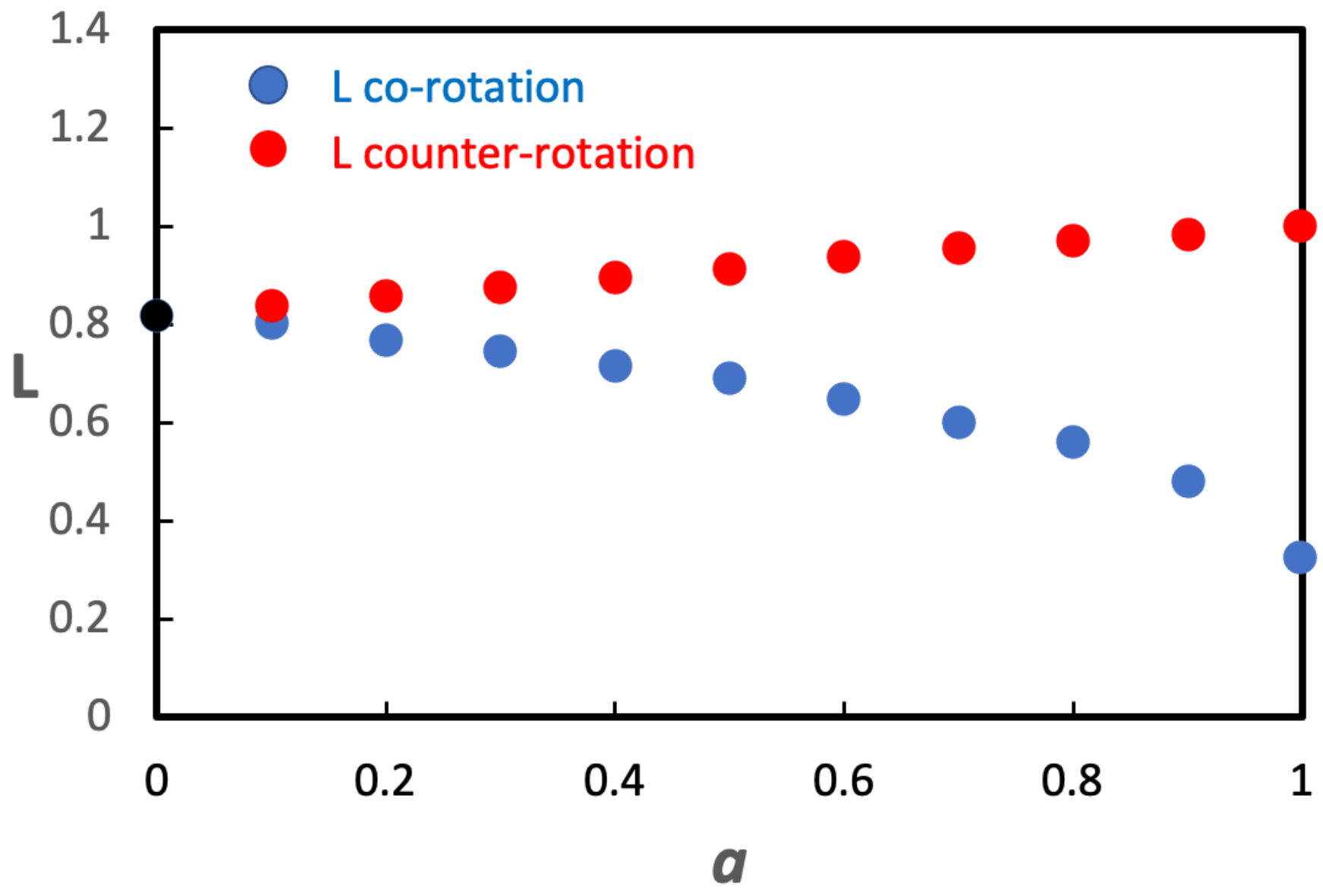

Figure 2: Angular momentum at $r_{ms}$ normalized to the angular momentum at $r_{ms}$ for a high spinning black hole accreting in counter-rotation, as a function of dimensionless black hole spin. Red and blue curves converge to the same angular momentum at zero spin as indicated by the data point in black.

In Figure 2 we show how the angular momentum of gas that accretes onto the black hole from the marginally stable orbit depends on the value of black hole spin. We scale or normalize the angular momentum to the angular momentum at the marginally stable orbit for a black hole spinning at a = 0.998 surrounded by an accretion disk in counter-rotation. Figure 2 allows one to appreciate why spinning a high spinning black hole down takes about an order of magnitude less time than it does to spin a zero spinning black hole up to high spin, at a given accretion rate.

As gas accretes onto the black hole from $r_{ms}$, both black hole spin and black hole mass change. The change in the black hole mass also depends on the location of $r_{ms}$ and can be obtained by evaluating the distribution of energy as a function of radius, i.e. energy counterparts to equations (1) and (2) above. One finds the black hole to gain an amount of mass given by (Raine & Thomas 2009)

$$\Delta m = \int dm(1 - 2m/3r_{ms})^{-0.5} \qquad (16)$$

where m is the mass of the black hole and the added mass $\Delta m$ is given as an expression with both Newton's constant and the speed of light equal to unity. As accretion proceeds, the marginally stable orbit decreases, acquires the value $r_{ms} = 6r_g$ when the black hole stops rotating, and then decreases further as it spins up via a co-rotating accretion disk. Our interest is only in the time to spin the black hole down to near zero spin. The crucial element to determine the timescale for spin down is the accretion rate. For a given accretion rate f, the time to build the mass by Δm is given in equation (18). If the accretion rate is constant, the time is given by equation (19).

$$dm/dt = f \qquad (17)$$

$$T = \int dm/f. \qquad (18)$$

$$T = \Delta m/f\ . \qquad (19)$$

For an accretion rate that is the Eddington value, a rapidly spinning counter-rotating black hole spins down to zero spin in just under 8 x $10^6$ years. Therefore, an FRII jet lives no longer than 8 x $10^6$ years if fed at the Eddington limit. But an FRII HERG does not need to be accreting at the Eddington limit. It could accrete at 10% the Eddington limit and still be a HERG. In this case, it would spin down to zero spin in 8 x $10^7$ years. For lower accretion rates, the timescale for spin down increases. If the FRII jet is powerful enough, it produces a strong feedback effect on the accretion flow, lowering the accretion rate, and allowing the FRII jet phase to last longer. In such cases of powerful jet feedback, the FRII jet affects the structure of the accretion disk, forcing it to evolve into an advection dominated accretion flow (ADAF). The boundary between a thin disk and an ADAF is prescribed from theory to be at $10^{-2}$ the Eddington accretion rate. Hence, at 5 x $10^{-2}$ the Eddington accretion rate, the object may still be characterized as an FRII HERG and the jet lifetime would increase to 1.6 x $10^8$ years. We should also note that jet lifetimes are effectively limited by some threshold low spin value below which the jet may no longer be classified as an FRII if even visible. From theory, we can estimate this to be below a spin value of about 0.1 but with some uncertainty that would also depend on the environment. Overall, we can estimate that an FRII HERG lives at most about $10^8$ years. This value appears as a red diamond on the right hand side of Figure 1.

Since the transition in cooling is not abrupt (e.g. Giustini & Proga 2019), the transition from cold mode accretion into an ADAF is also gradual, and the model prescribes accretion rates to barely cross the boundary into an ADAF during counter-rotation. In other words, counter-rotation may have ADAF accretion, but it has only recently entered the ADAF regime and accretion rates will therefore have values near $10^{-2}$ the Eddington accretion rate (see Garofalo, Evans & Sambruna 2010 for details). If one assumes such a value for the accretion rate in equation (17), one obtains a timescale for spin down of 8 x $10^8$ years. But, as mentioned above, the initial state is a near-Eddington accreting black hole so the timescale prescribed in the model for systems that evolve into ADAF accretion, must be lower. In short, the black hole spins halfway down from maximal spin at near the Eddington accretion rate (as an FRII HERG) which takes 4 x $10^6$ years, followed by accretion at $10^{-2}$ the Eddington accretion rate (as an FRII LERG), requiring 4 x $10^8$ more years to spin the black hole down to zero spin. Therefore, the model prescription for FRII lifetimes spans the range 8 x $10^6$ to 4 x $10^6$ + 4 x $10^8$ years which is roughly 4 x $10^8$ years and is our second theoretical data point in Figure 1 (the blue diamond). Because these lifetimes occur in systems with strongest FRII jet feedback, and the model prescribes the strongest jet feedback to occur more in denser environments, there is a model-prescribed environment-dependence to the maximum FRII jet lifetimes that is worth mentioning although it is not the focus of this work.

Although our focus has not been on giant radio quasars and giant radio galaxies, such objects serve an important role as guideposts, allowing us to understand better the constraints on radio quasars and radio galaxies. As re-triggered counter-rotating black holes, giant radio quasars have the opportunity to generate jets that extend beyond the kiloparsec lengths reached by their radio galaxy ancestors, and to experience longer lived FRII phases (Garofalo 2022). Accordingly, it is interesting to note that while giant radio quasars and giant radio galaxies exceed the prescribed theoretical maximum lifetimes for radio quasars (the red diamond in Figure 1), this is not true for radio quasars. If counter-rotation is not relevant to FRII jets, there is no reason for the latter to be limited in this way. In addition, if the evolution in time from quasar mode to ADAF mode (as the model prescribes) is not the way such systems change over time, there is no reason for both regular radio quasars and radio galaxies as well as their giant counterparts to have blue objects experiencing longer jet lifetimes than red ones. But it is instead required from theory.

## 4. Conclusion

The timescales for FRII systems obtained from theory are tantalizingly compatible with those inferred from the data as seen from the diamonds added on the right-hand side of Figure 1 to represent theoretical timescales. While spin-down timescales are constrained by the rate of accretion, this same constraint on time for FRI jets in the theory is rather weak because a black hole that feeds forever simply remains a high spinning co-rotating black hole. The real constraint, instead, is the amount of fuel. We have not gone into model details but FRI systems are late stages in the evolution of radio galaxies that were once FRII. Their accretion rates continue to drop over time and can be orders of magnitude lower than the Eddington accretion rate. As a result, FRI systems accreting in ADAF can last characteristic timescales that are on the order of the age of the universe, making them effectively unconstrained in time. In closing, we highlight that jet dynamical lifetimes are characterized by large uncertainties (e.g. Wojtowicz et al 2021) and that until these have been sufficiently reduced, a clear picture cannot emerge and caution should be exercised in comparing jet lifetime from the model with dynamical timescales from data. Nonetheless, if FRII lifetimes can robustly be shown to be limited to within a half billion years, it would constitute strong evidence for counter-rotating black holes in active galaxies and the evidence appears to be pointing in that direction.

While constraints on the lifetimes of FRII jetted AGN have existed for a decade or so, data has recently emerged to solidify the case that FRII and FRI jets are different in some fundamental way. We have argued over the last decade that opening the counter-rotating window for black hole accretion allows many disparate observations to come together under a simple evolutionary picture that at its heart explains the radio loud/radio quiet dichotomy. In this work we highlight the otherwise coincidental match between the lifetimes of FRII jets in quasars and radio galaxies, showing how to understand the difference in the evolution of FRII jets compared to FRI jets.

Unlike FRI LERG whose jet lifetimes are effectively unconstrained, FRII jets in either LERG or HERG form, are limited to lifetimes within hundreds of millions of years due to accretion spinning black holes down, a process that is time-limited in a way that spinning black holes up is not.

*Acknowledgements.* I thank Dr. Marek Jamrozy and Dr. Dorota Koziel-Wierzbowska for sharing their expertise. In addition, I acknowledge the role of 4 referees at FrASS but thank referees 3 and 4 for pointing to the need for clarification on key points. The reason FRIIs appear to prefer less dense environments compared to FRIs is that they live longer in such environments. This resolves an interesting point raised by referee 4 that was not included in the paper because it is outside its scope. The issue is discussed in our work on X-shaped radio galaxies in 2020.